\documentclass[aps,prl,twocolumn,showpacs,preprintnumbers,amsmath,aps]{revtex4}
\usepackage{graphicx,hyperref}
\usepackage{bm}
\usepackage{subfigure}

\def\nbC{{\mathchoice {\setbox0=\hbox{$\displaystyle\rm C$}%
\hbox{\hbox to0pt{\kern0.4\wd0\vrule height0.9\ht0\hss}\box0}}
{\setbox0=\hbox{$\textstyle\rm C$}\hbox{\hbox
to0pt{\kern0.4\wd0\vrule height0.9\ht0\hss}\box0}}
{\setbox0=\hbox{$\scriptstyle\rm C$}\hbox{\hbox
to0pt{\kern0.4\wd0\vrule height0.9\ht0\hss}\box0}}
{\setbox0=\hbox{$\scriptscriptstyle\rm C$}\hbox{\hbox
to0pt{\kern0.4\wd0\vrule height0.9\ht0\hss}\box0}}}}
\def\nbQ{{\mathchoice {\setbox0=\hbox{$\displaystyle\rm
Q$}\hbox{\raise
0.15\ht0\hbox to0pt{\kern0.4\wd0\vrule height0.8\ht0\hss}\box0}}
{\setbox0=\hbox{$\textstyle\rm Q$}\hbox{\raise
0.15\ht0\hbox to0pt{\kern0.4\wd0\vrule height0.8\ht0\hss}\box0}}
{\setbox0=\hbox{$\scriptstyle\rm Q$}\hbox{\raise
0.15\ht0\hbox to0pt{\kern0.4\wd0\vrule height0.7\ht0\hss}\box0}}
{\setbox0=\hbox{$\scriptscriptstyle\rm Q$}\hbox{\raise
0.15\ht0\hbox to0pt{\kern0.4\wd0\vrule height0.7\ht0\hss}\box0}}}}
\def\nbT{{\mathchoice {\setbox0=\hbox{$\displaystyle\rm
T$}\hbox{\hbox to0pt{\kern0.3\wd0\vrule height0.9\ht0\hss}\box0}}
{\setbox0=\hbox{$\textstyle\rm T$}\hbox{\hbox
to0pt{\kern0.3\wd0\vrule height0.9\ht0\hss}\box0}}
{\setbox0=\hbox{$\scriptstyle\rm T$}\hbox{\hbox
to0pt{\kern0.3\wd0\vrule height0.9\ht0\hss}\box0}}
{\setbox0=\hbox{$\scriptscriptstyle\rm T$}\hbox{\hbox
to0pt{\kern0.3\wd0\vrule height0.9\ht0\hss}\box0}}}}
\def\nbS{{\mathchoice
{\setbox0=\hbox{$\displaystyle     \rm S$}\hbox{\raise0.5\ht0%
\hbox to0pt{\kern0.35\wd0\vrule height0.45\ht0\hss}\hbox
to0pt{\kern0.55\wd0\vrule height0.5\ht0\hss}\box0}}
{\setbox0=\hbox{$\textstyle        \rm S$}\hbox{\raise0.5\ht0%
\hbox to0pt{\kern0.35\wd0\vrule height0.45\ht0\hss}\hbox
to0pt{\kern0.55\wd0\vrule height0.5\ht0\hss}\box0}}
{\setbox0=\hbox{$\scriptstyle      \rm S$}\hbox{\raise0.5\ht0%
\hboxto0pt{\kern0.35\wd0\vrule height0.45\ht0\hss}\raise0.05\ht0%
\hbox to0pt{\kern0.5\wd0\vrule height0.45\ht0\hss}\box0}}
{\setbox0=\hbox{$\scriptscriptstyle\rm S$}\hbox{\raise0.5\ht0%
\hboxto0pt{\kern0.4\wd0\vrule height0.45\ht0\hss}\raise0.05\ht0%
\hbox to0pt{\kern0.55\wd0\vrule height0.45\ht0\hss}\box0}}}}
\def\nbZ{{\mathchoice {\hbox{$\sf\textstyle Z\kern-0.4em Z$}}
{\hbox{$\sf\textstyle Z\kern-0.4em Z$}}
{\hbox{$\sf\scriptstyle Z\kern-0.3em Z$}}
{\hbox{$\sf\scriptscriptstyle Z\kern-0.2em Z$}}}}

\begin{document}

\title{Renormalization group analysis of the random first order transition}

\author{Chiara Cammarota} \email{chiara.cammarota@cea.fr}
\affiliation{IPhT, CEA/DSM-CNRS/URA 2306, CEA Saclay, F-91191 Gif-sur-Yvette Cedex, France}

\author{Giulio Biroli} \email{giulio.biroli@cea.fr}
\affiliation{IPhT, CEA/DSM-CNRS/URA 2306, CEA Saclay, F-91191 Gif-sur-Yvette Cedex, France}

\author{Marco Tarzia} \email{tarzia@lptmc.jussieu.fr}
\affiliation{LPTMC, CNRS-UMR 7600, Universit\'e Pierre et Marie Curie,
bo\^ite 121, 4 Pl. Jussieu, 75252 Paris c\'edex 05, France}

\author{Gilles Tarjus} \email{tarjus@lptmc.jussieu.fr}
\affiliation{LPTMC, CNRS-UMR 7600, Universit\'e Pierre et Marie Curie,
bo\^ite 121, 4 Pl. Jussieu, 75252 Paris c\'edex 05, France}

\date{\today}

\begin{abstract}

We consider the approach describing glass formation in liquids as a progressive trapping in an exponentially large number of metastable states. To go beyond the mean-field setting, we provide a real-space renormalization group (RG) analysis of the associated replica free-energy functional. The present approximation yields in finite dimensions an ideal glass transition similar to that found in mean field. However, we find that along the RG flow the properties associated with metastable glassy states, such as the configurational entropy, are only defined up to a characteristic length scale that diverges as one approaches the ideal glass transition. The critical exponents characterizing the vicinity of the transition are the usual ones associated with a first-order discontinuity fixed point.
\end{abstract}


\maketitle

In the ongoing search for a general theory of glass formation, the random first-order transition (RFOT) approach has proven to be a strong candidate, establishing what appears to be an intricate  mean-field (MF) description of supercooled liquids and glasses.\cite{KTW89,lubchenko07,BBreview} This MF treatment predicts a scenario with two critical temperatures $T_d$ and $T_K$, the upper one $T_d$ being a dynamical singularity akin to the mode-coupling transition and the lower one $T_K$ a thermodynamic ideal glass transition characterized by a vanishing of the configurational entropy associated with the logarithm of the number of metastable states. This scenario has received support from MF-like calculations on glassforming liquid models.\cite{wolynesDFT,franz-parisi,mezard-parisi,zamponi,mezard10} 
The RFOT theory assumes that this MF picture retains some validity in finite-dimensional systems and proceeds by accounting for ergodicity restoring and disappearance of metastability in supercooled liquids between $T_d$ and $T_K$ through an entropy-driven nucleation process coupled with a mosaic view of the liquid configurations.\cite{KTW89,lubchenko07,BBreview,BB04} Testing the validity of this appealing but still fragile scenario is of major interest. In addition to computer simulations of model liquids,\cite{cavagna-PS,cavagna-exponents} analytical work has so far been done in two directions, taking the MF result as a starting point: instanton calculations for the escape from glassy metastable states\cite{franz-instanton,dzero} and studies of disordered models with long-range interactions in the Kac limit.\cite{franz-instanton,franz-kac} The latter in particular have shown that both static and dynamic  correlation lengths can be computed and that the MF infinite-range limit is not singular. Going beyond these approaches however requires a renormalization group (RG) treatment.\\
We provide in this letter the first steps towards an RG treatment of glass formation beyond the RFOT MF theory. To this end, we consider the Migdal-Kadanoff (MK) real-space RG of a Ginzburg-Landau model which is commonly taken to be in the ``universality class'' of structural glass-formers as it generically displays the two-temperature scenario at the MF level. We use the replica formalism and, for convenience, we use hierarchical lattices, on which the MKRG is known to be exact.\cite{berker79}\\
Our starting point is the replica MF theory of structural glasses, in which one studies the distribution of putative metastable glassy states by introducing $m-1$ copies (or replicas) of the same liquid system coupled with a small attractive interaction whose amplitude is set to zero after taking the thermodynamic limit.\cite{monasson95,mezard10} By keeping the leading terms in the local order parameter, which is the similarity or ``overlap'' between different states, one obtains the following Ginzburg-Landau functional\cite{dzero}
\begin{equation}
\label{eq_GLfunctional}
S\left[ \mathbf q \right] =  \int d^d x \bigg\{\frac{1}{2} \sum_{a,b=1}^m  \left( \partial q_{ab}(x)\right)^2 +  V(\mathbf{q}(x)) \bigg\}
\end{equation}
with
\begin{equation}
\begin{split}
\label{eq_replica_potential}
V = \sum_{a,b=1}^m ( \frac{t}{2}q_{ab}^2 -\frac{u+w}{3}q_{ab}^3 + \frac{y}{4}q_{ab}^4 )   - \frac{u}{3}\sum_{a,b,c=1}^mq_{ab}q_{bc}q_{ca}
\end{split}
\end{equation}
where $\mathbf q$ denotes the set of elements $\{q_{ab}\}$ (by construction, $q_{aa}=0$) and the overlap $q_{ab}(x)$ is physically associated with a local Debye-Waller factor characterizing molecular motion in the glass-forming liquid\cite{dzero}; for simplicity, the only temperature dependence is taken in $t= \frac{T-T_0}{T_0}$, with $T_0$ setting the temperature scale, while $u,w,y >0$ are considered as independent of temperature. This ``real replica''  method allows one to 
obtain the properties of the metastable states from the knowledge of the replica partition function, $\mathcal Z(m) = \int \prod_{ab}\mathcal D q_{ab}(x) \exp(-S[\mathbf q])$.
 The mean free energy of a typical equilibrium state and the corresponding configurational entropy read respectively $\beta F= - \partial \log \mathcal Z(m)/\partial m$ and $S_c=-m^2\partial (\log \mathcal Z(m)/m) /\partial m$. The number $m$ of replicas should be analytically continued to $1$ in the equilibrium liquid phase and to a value less than one in the ideal glass phase, if present.\cite{mezard10} At the MF level, \textit{i.e.} by looking for the uniform saddle-points of Eq.~(\ref{eq_GLfunctional}), one finds that the order parameter $q_{ab}$ is zero above a temperature $T_d$, such that $t_d=\frac{w^2}{4y}$, and that below $T_d$ appears another uniform solution with a  replica symmetric (RS) structure $q_{ab}= q>0$ for $a\neq b$. This solution, when plugged into the expression for $F$ and $S_c$, yields the properties 
of the metastable glassy states. Finally, at a temperature $T_K$ such that $t_K=\frac{2w^2}{9y}$, there is a RFOT with a coexistence between a zero-overlap phase and a high-overlap one, transition with zero latent heat and vanishing configurational entropy density. Below $T_K$,  the system is in an ideal glass phase characterized by a nonzero overlap matrix and a value of $m$ less than $1$.\\ 
To go beyond MF, we consider a real-space MKRG approach, which becomes exact on hierarchical diamond-like lattices, and apply it to a lattice version of the effective Hamiltonian in Eq.~(\ref{eq_GLfunctional}). Such lattices are built iteratively by replacing each bond between sites by a fixed number of new bonds which, to mimic Euclidean $d$-dimensional lattices, is taken equal to $2^d$. After $n$ iterations, the volume of the system, which is equal to the total number of original bonds, is equal to $2^{nd}$ whereas the ``distance'' between the boundary sites is equal to $2^n$ bonds: this naturally fixes the length scale after $n$ iterations as $\ell_n=2^n$. The procedure is illustrated in the inset of Fig.~1. The main advantage of this RG procedure is that the renormalized effective pair interaction between two sites at the $n-$th step of renormalization, $W_n(\mathbf q^1,\mathbf q^2)$, satisfies a closed equation written in terms of the pair interaction $W_{n-1}(\mathbf q^1,\mathbf q^2)$:
\begin{equation}
\begin{split}
\label{eq_integral_recursion}
&2^{-(d-1)}W_{n}(\mathbf q^1 ,\mathbf q^2)=  \\&\log \int \prod_{a,b}dq_{ab} \exp\bigg\{W_{n-1}(\mathbf q^1,\mathbf q )+  V(\mathbf q)+  W_{n-1}(\mathbf q ,\mathbf q^2)\bigg \}
\end{split}
\end{equation}
where the labels $1$ and $2$ denote the value of two renormalized sites from which emanate $2^{n(d-1)}$ original bonds. 
At the $n$th iteration, the original lattice is replaced by a renormalized one where the unit length is $\ell_n$ and 
the pair interaction between sites is $W_n(\mathbf q^1,\mathbf q^2)$. 
A remaining obstacle is that even if the integration in Eq.~(\ref{eq_integral_recursion}) is purely local, it involves the components of a general $m\times m$ replica matrix with $m$ continued to real values, and there is no general solution to this problem. However, the physics we aim at describing is related to that of first-order-like transitions; as a result, $W_n$ for generic $\mathbf q^1$ and $\mathbf q^2$ is expected to grow rapidly with the number of iterations,  as $2^{nd}$ or $2^{n(d-1)}$, so that when $d\geq 2$ the term in the exponential of the left-hand side of Eq.~(\ref{eq_integral_recursion}) becomes very large after a few iterations only. In consequence, we approximate the full integral by a steepest-descent calculation. 
In addition, guided by the MF solution, we looked for $p$-step RSB saddle-points. Since we found that they all reduce 
to the RS saddle point, in the following we shall only focus on the latter. All these restrictions and limitations will be further discussed below.\\
In the RS case, any matrix $\mathbf{q}$ is characterized by a single parameter $q$ and, after introducing $\widetilde W_{n}(q_1,q_2)=\frac{W_{n}(q_1,q_2)}{m-1}$ and $\widetilde V(q)=\frac{V(q)}{m-1}$, the iteration equation simplifies to
\begin{equation}
\label{eq_min_recursion}
\begin{split}
&\widetilde W_{n}(q_1,q_2)=\\& 2^{d-1}\min_{q}\left\lbrace \widetilde W_{n-1}(q_1,q)+\widetilde W_{n-1}(q,q_2) + \widetilde V(q) |_{m=1}\right\rbrace 
\end{split}
\end{equation}
where $\widetilde V(q)= \frac{t}{2}q^2-\frac{w}{3}q^3-(m-1)\frac{u}{3}q^3+\frac{y}{4}q^4$ has a unique minimum at $q=0$ above $T_d$ and an additional metastable minimum at $q=q^*(T)=\frac{1}{2y}(w+\sqrt{w^2-4yt})$ between $T_d$ and $T_K$, minimum which becomes the deepest one below $T_K$(the bare parameters are arbitrarily chosen such that $t_K=q(t_K)=1$). At the start of the RG flow, the ``bare interaction'' is taken equal to $\widetilde W_0(q_1,q_2)=(q_1-q_2)^2/2$. It is easily derived from Eq.~(\ref{eq_min_recursion}) that $\widetilde W_{n}(0,0)$ does not flow and remains equal to zero (and so does the free energy $F_n(0,0)$).
\begin{figure}
  \centering
  \includegraphics[angle=0, width=0.47\textwidth]{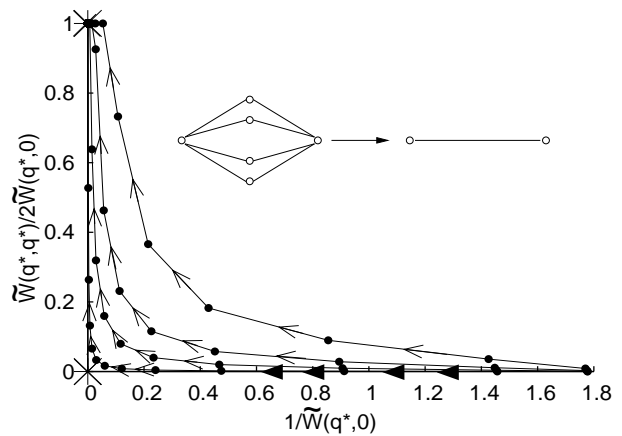}
\caption{RG flow on the hierarchical lattice with $d=3$ for several initial conditions parametrized by $T$($=1.01, 1.003, 1.001, 1.0001$ from top to bottom): parametric plot of $\widetilde W_{n}(q^*,q^*)/(2\widetilde W_{n}(q^*,0))$ versus $1/\widetilde W_{n}(q^*,0)$. The two fixed points on the y-axis correspond to the RFOT ($y=0$) and to the ``normal''  liquid ($y=1$). Inset: Elementary step illustrating the RG on a hierachical lattice corresponding to $d=3$.}
  \label{RGflow}
\end{figure}
\\We first consider the liquid between $T_d$ and $T_K$. Asymptotically, \textit{i.e.} for large $n$, the system flows to a trivial disordered fixed point corresponding to a ``normal'' liquid and uncoupled replica.
\setlength{\unitlength}{0.2000bp}
\begin{figure}[h!]
  \centering
 \includegraphics[angle=-90,width=0.47\textwidth]{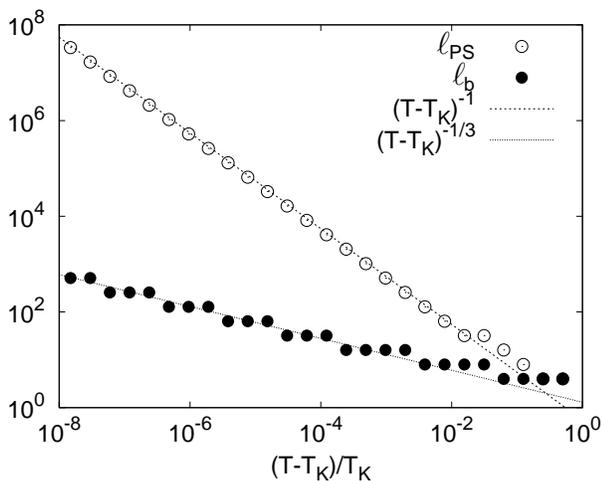}
\put(-351,-110){\large $\ell$}
\put(-322,-166){\large $\ell$}
\caption{Temperature dependence of the point-to-set correlation length $\ell_{PS}$ and of the penetration length $\ell_b$ in the liquid phase above $T_K$ for a hierarchical lattice with $d=3$.}
  \label{PSlb}
\end{figure}
In this case, at each RG step, the minimum in Eq.~(\ref{eq_min_recursion}) is in $q=0$, which leads for large enough $n$ to $\widetilde W_n(q_1,q_2)\simeq \widetilde W_n(q_1,0) + \widetilde W_n(q_2,0)$ for $q_1, q_2\neq 0$:  boundary conditions then have no influence on the bulk. 
This behavior takes place along the RG flow beyond a length scale $\ell_{PS}$ which depends on the initial condition, here parametrized by $T$. More precisely, we define $\ell_{PS}=2^{n_{PS}}$ from the number $n_{PS}$ of iterations needed to make the value of $q$ corresponding to the minimum in Eq.~(\ref{eq_min_recursion}) for both boundaries fixed in the metastable state $q^*$ drop below the value $q^*/2$ (beyond this scale, the minimum is found in $q=0$ for all boundary conditions). From its definition, $\ell_{PS}$ corresponds to a ``point-to-set'' correlation length:\cite{BB04,montanari-semerj} above $\ell_{PS}$, the boundary conditions in the metastable glassy state do not affect the deep interior of the liquid whereas below $\ell_{PS}$, they determine the state of the liquid. Exactly at $T_K$, one finds that the minimum sticks at $q^*$ when the boundary conditions are fixed at $q^*$: then, $\widetilde W_{n}(q^*,q^*)$ stays equal to zero, just as $\widetilde W_{n}(0,0)$. This corresponds to a first-order transition (a RFOT) with a coexistence between a liquid phase with $q=0$ and an ideal glass phase with $q=q^*(T_K)$. The fact that the value of $T_K$ is itself not renormalized is a consequence of the minimization procedure (see also below).\\
In Fig.~1, we illustrate the RG flow for several initial conditions in the case of a hierarchical lattice mimicking a $3$-dimensional system (shown in Fig.~1). The behavior is strongly reminiscent of that observed in a conventional first-order transition,\cite{klein81} with the point-to-set length playing the role of the scale above which the low-$T$ metastable phase disappears. We plot in Fig.~2 the $T$-dependence of $\ell_{PS}$. It follows a power law, $\ell_{PS}(T) \sim (T-T_K)^{-1}$.
Interestingly, there appears to be another characteristic length scale $\ell_b$, which we call the ``penetration length'' as it describes how far the ``amorphous order'' fixed by the metastable boundary condition penetrates in the liquid; as such,  $\ell_b$  seems
to be related to the pattern repetition length introduced in \cite{kurchan}. Specifically, we compute it from the number of iterations $n_b$ required to make the value of $q$ corresponding to the minimum in the RG equation for $\widetilde W_{n}(q^*,0)$ drop below $q^*/4$. Above this length, the minimum is found at $q=0$. The dependence of $\ell_b$ on initial conditions is shown in Fig.~2: it goes as $(T-T_K)^{-\frac{1}{d}}$ .
While this dependence might be specific to the structure of the hierarchical lattice, it is noteworthy that the exponent $1/d$ is 
the same one obtained in \cite{kurchan} for the pattern repetition length and also coincides with the standard exponent for the ``persistence length'' near a first-order discontinuity fixed point.\cite{fisher82} Note that the asymptotic behavior of $\ell_{PS}$ and $\ell_b$ close to $T_K$ can be also derived analytically (details will be presented elsewhere).\\
In order to obtain a description of the renormalized liquid on the scale $\ell_n$, it is useful to define and compute at the $n$th
RG iteration both the configurational entropy, $S_c(\ell_n,T)={\beta[F_n(q^*,q^*)-F_n(0,0)]}$, and the interface free-energy  
$\Upsilon(\ell_n,T)=F_n(q^*,0)-[F_n(q^*,q^*)-F_n(0,0)]/2$ between the liquid phase with $q=0$ and the glass phase with $q=q^*$.
Even though close to $T_K$ $S_c$ is very small on microscopic scales, it increases by RG transformations as $\ell_n^{d}$ whereas $\Upsilon$ grows as $\ell_n^{d-1}$ only. In consequence, $S_c$ and $\Upsilon$ eventually become of the same order. This happens on a scale of order $\ell_{PS}$. At this point, the free-energy gain of having $q=0$ onsite becomes overwhelming compared to the 
free-energy cost due to a mismatch between the overlap values; ; the renormalized value of Upsilon then drops to zero, see Fig. 3, and the MF quantity Sc is no longer well-defined. \cite{BB04} Note that on the lengthscale just below $\ell_{PS}$ the configurational entropy density $s_c=S_c/\ell_{n}^d$  linearly approaches zero as $T-T_K$, whereas the surface tension  $\upsilon=\Upsilon/\ell_{n}^{d-1}$ remains nonzero at the transition, see Fig. 3.\footnote{The renormalized surface tension should not be confused with the microscopic surface tension between two different metastable states. These are two different quantities. The latter has been recently studied numerically in \cite{cavagna-exponents} and analytically in \cite{franzst}.}
 \begin{figure}
\setlength{\unitlength}{0.2000bp}
  \centering
   \includegraphics[angle=-90,width=0.47\textwidth]{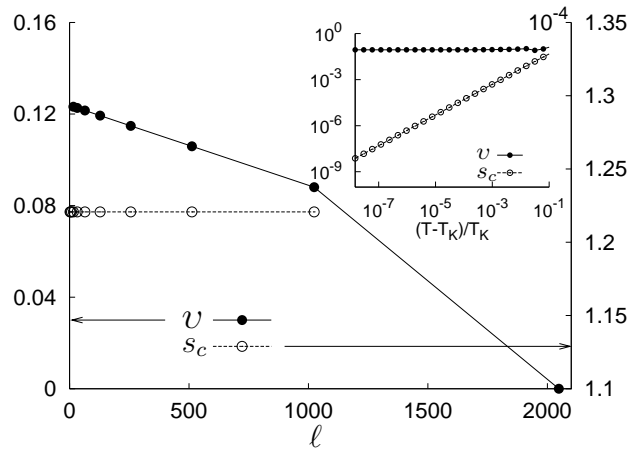}
\put(-303,-308){$\upsilon$}
\put(-303,-343){$s_c$}
\put(-858,-625){\Large $\upsilon$}
\put(-858,-670){\large $s_c$}
\put(-618,-850){\large $\ell$}
\caption{Renormalized configurational entropy density $s_c$ and surface tension $\upsilon$ for a hierarchical lattice with $d=3$. Main plot: Evolution with the RG scale $\ell_n$ for $T=T_K+0.5^{12}$. Inset: $T$-dependence of the values of $s_c$ and $\upsilon$ just before the point-to-set scale.}
  \label{Sc}
\end{figure}
The picture we obtain is that after $n_{PS}$ RG iterations, the renormalized system is like a liquid at its ``onset temperature'' where the (renormalized) PS length is equal to one. The difference with a normal liquid are the values of $\Upsilon$ and $S_c$, which are very large 
compared to $T$, {\it i.e.} the liquid is at very low $T$ compared to the typical scale of the interaction. 
This--naively--suggests that the relaxation time could be obtained by assuming an Arrhenius law
at the scale $\ell_{PS}$: $\log \tau \propto \Upsilon/T \propto 1/(T-T_K)^{d-1}$.\\
We now briefly discuss the situation below $T_K$. The calculation is similar to that performed above $T_K$ except that when describing the equilibrium ideal glass phase with $q=q^*(T)$, the replica  parameter $m$ should also be optimized, which leads to  $m_{eq}(T)<1$,\cite{mezard10}  and that at the MF level, it is the liquid phase with $q=0$ which is metastable. We again find that the $p$-step RSB saddle-points reduce to the RS one (since $m<1$, the extremization on $q$ now becomes a maximization\cite{mezard10}). We define a point-to-set correlation length by studying when the minimum involved in the iteration equation for $\widetilde W_{n}(0,0)$ increases above $q^*/2$. We find that the equilibrium point-to-set correlation length defined for $m=m_{eq}(T)$ diverges as one approaches $T_K$ from below with the same exponent as from above, \textit{i.e.} as $(T_K-T)^{-1}$.\\
From the above results, it appears that the physics in the vicinity of a RFOT to an ideal glass is controlled by a first-order discontinuity fixed point with standard exponents. Although the resulting physical picture is similar to the phenomenological
one put forward in \cite{KTW89,BB04}, the values of the exponents differs from \cite{KTW89}.
Recent computer simulations of an atomic glass-forming liquid model,\cite{cavagna-exponents} find that some exponents are indeed standard, but others are not. It is therefore worth discussing the limitations of the present RG analysis. The starting replica Ginzburg-Landau model and the choice of hierachical lattices could of course be criticized, but the main potential shortcoming of the present study is the steepest-descent approximation. We have argued that the latter is justified after a few iteration steps because  the factor in the Boltzmann weight grows rapidly. In particular, if one starts with an arbitrarily strong random first-order transition at the MF level (this may not correspond to any actual glass-forming liquid but could nonetheless provide an interesting limiting case) and consider an initial condition deep in the glass phase, it is unlikely that the  latter and the associated random first-order transition would be destroyed by the few first iterations before minimization becomes justified. However, the parameters of the theory, such as the value of $T_K$, would clearly be renormalized by these first steps, so that minimization would start with effective parameters in place of the bare ones.\footnote{Note that some renormalization of parameters can be obtained in a simple but not fully consistent way by either replacing the minimization by an integration over a single variable $q$, which for instance is enough to destroy the transition in $d=1$, or computing the Gaussian fluctuations around the RS saddle-point.}
Furthermore, fluctuations may actually play an important role in determining the exponents of $\ell_b$ and consequently $\ell_{PS}$ since these involve the flow of $S_c$ that grows as $(T-T_K)\ell_n^d$ and remains small until the scale $\ell_b$. Another questionable aspect of our minimization scheme is the choice of saddle-points among RS and $p$-step RSB solutions only. Could we be missing relevant minima ? A way to get around the above problems would be to somewhat invert the replica treatment and infer from a replicated free-energy functional with a strong RFOT at the MF level an effective model with quenched disorder that could be directly studied by the MKRG on hierarchical lattices. This is a promising but arduous route, as there are at present no known disordered models displaying remains of the two-temperature scenario in finite dimensions.\cite{RMB}  (See however the recent proposal of a hierarchical random energy model in \cite{parisi10} and numerical simulations on a Kac version of the random orthogonal model \cite{sarlat}.)
This nonetheless appears as the only means to check if, in a full-blown RG analysis, the trivial RFOT discontinuity fixed point found here is confirmed, replaced by a nontrivial one, or else if the RFOT itself is completely suppressed.\\
  We would like to thank J.-P. Bouchaud, S. Franz, J. Kurchan and M. M\'ezard for useful discussions and acknowledge 
  partial financial support from ANR DYNHET.

\end{document}